\documentclass{article}
\usepackage{amsmath}

\input{tcilatex}

\begin{document}

\title{Evolution of the D3-brane for dynamical embeddings }
\author{Pawel Gusin, Jerzy Warczewski \\
%EndAName
University of Silesia, Institute of Physics, ul. Uniwersytecka 4, \\
PL-40007 Katowice, Poland}
\maketitle

\begin{abstract}
The Dirac-Born-Infeld (DBI) action of the 3-dimensional brane for its
dynamical embeddings and the gauge fields has been studied. The evolution of
both the D3-brane and the ambient space has been obtained. For the special
constraint put on the transverse coordinates a family of the contracting and
expanding spaces has been found.

\begin{description}
\item[PACS] : 11.25.-Uv; 11.27.+d

\item[Keywords] : DBI action, supersymmetry
\end{description}
\end{abstract}

\section{Introduction}

In the presence of D-branes supersymmetry of the background is broken. The
survived supersymmetries are given by the Killing spinors. If these spinors
are projected onto the D-brane, then one obtains as a result an equation for
the survived supersymmetry generators [1]. The number of the unbroken
supersymmetries is equal to the dimension of the solution space of this
equation. The supresymmetry charges are minimal in the case of the BPS
states. It means that masses are equal to the gauge charges. Thus the
Hamiltonian (the energy of the system) takes the special form. For the D$%
_{p} $-branes described by the DBI action, relations between the embeddings,
gauge fields and geometric and topological properties of the ambient space
are discussed in [2]. This is related to the concept of a calibrated
submanifold [3]. The calibrated submanifold minimizes the DBI action. Thus
the calibration bound gives the BPS bound. In the present paper evolution of
both a D3-brane and the ambient space has been investigated. This evolution
is induced by the deformed metric which is related to the vanishing of the
DBI Lagrangian. In Section 2 we recall the form of the DBI action and
constraints which appear in this action in the case when both the gauge
fields and embeddings are dynamic. Since the Lagrangian for DBI action is
invariant under diffeomorphism the Hamiltonian for the DBI action is just
equal to the sum of constraints. In Section 3 we present the case of a
dynamic embedding, which means that both the coordinates transverse to the
brane and the gauge fields depend on time. We put an isotropic constraint on
the transverse velocities. This constraint depends on two parameters. In
this case the deformed metric leads to the evolution both of the D-brane and
the ambient space. For the different parameters we are going to obtain
different evolutions of the D-brane and the ambient space. For the special
values of these parameters the de Sitter space is obtained.

\section{DBI Lagrangian and the constraints}

The low energetic action in the flat ambient space-time for a $Dp$-brane is
given by the expression: 
\begin{eqnarray}
S &=&-T_{p}\int_{\mathcal{M}_{p+1}}e^{-\phi }\left( -\det \left( \gamma
_{\alpha \beta }+2\pi \alpha ^{\prime }F_{\alpha \beta }+B_{\alpha \beta
}\right) \right) ^{1/2}d^{p+1}\xi +  \notag \\
&&T_{p}\int_{\mathcal{M}_{p+1}}\sum\limits_{i}C_{i}\wedge \exp \left( 2\pi
\alpha ^{\prime }F+B\right) ,  \TCItag{2.1}
\end{eqnarray}%
where: $\phi $ is a dilaton field, $\alpha ,\beta =0,1,...,p$ and $p$ is a
spatial dimension of a D$p$-brane. The metric $\gamma _{\alpha \beta }$ on
the worldvolume is induced by the background metric $g_{MN}$:%
\begin{equation*}
\gamma _{\alpha \beta }=g_{MN}\partial _{\alpha }X^{M}\partial _{\beta
}X^{N},
\end{equation*}%
$X^{M}$\ is the embedding of $\mathcal{M}_{p+1}$\ into the ambient spacetime:%
\begin{equation*}
X^{M}=\left( X^{\alpha },X^{a}\right) ,
\end{equation*}%
$a=1,...,9-p$. The RR fields are denoted as $C_{i}$, $F_{\alpha \beta }$ is
the abelian gauge field strenght on the brane and $B_{\alpha \beta }$ is the
pullback of the bacgkround NS 2-form $B$. The symbols $F$ and $B$ denote:%
\begin{eqnarray*}
F &=&F_{\alpha \beta }d\xi ^{\alpha }\wedge d\xi ^{\beta }, \\
B &=&B_{\alpha \beta }d\xi ^{\alpha }\wedge d\xi ^{\beta }.
\end{eqnarray*}%
In the last integral in Eq. (2.1)\ are picked up only those forms with
degree which is equal to the dimension of the D$p$-brane. In the case of the
non-flat backgrounds the action (2.1) is corrected by the non-linear terms
in the curvature forms, both of $M$\ and the ambient spacetime [4].

\ It is well-known that a gauge field $A$ with the strength $F$ produces $%
D\left( p-2\right) $-brane by the WZ action and the fundamental string by
coupling to the background NS antisymmetric field $B$. In this paper the
purely geometrical backgrounds have been taken into account only. Thus the
WZ action vanishes. The case with $p=3$ only has been considered. In [5, 6]
the problem of the stability of supertubes and branes has been discussed.

Thus the action (2.1) is reduced to the DBI action and takes the form: 
\begin{equation}
S=T_{3}\int_{\mathcal{M}_{4}}d^{4}\xi e^{-\phi }\sqrt{-\left( 1-\frac{1}{2}%
\gamma ^{\alpha \gamma }\gamma ^{\beta \delta }\mathcal{F}_{\alpha \delta }%
\mathcal{F}_{\beta \gamma }\right) \det \gamma -Pff^{2}\left( \mathcal{F}%
\right) },  \tag{2.2}
\end{equation}%
where $\mathcal{F}_{\alpha \beta }=2\pi \alpha ^{\prime }F_{\alpha \beta }$
(in the case considered $B=0$) and $Pff^{2}\left( \mathcal{F}\right) =\det
\left( \mathcal{F}\right) $. In the non-flat background cases one should add
to the action (2.2) non-linear terms in the curvature. The Lagrangian in the
Eq.(2.2) is expressed by the electric field $E_{m}=-F_{0m}$ and the magnetic
field $B_{m}=\frac{1}{2}\varepsilon _{mnp}F^{np}$ and assumes the following
form:%
\begin{equation*}
L=T_{3}e^{-\phi }\sqrt{-\left( 1+\left( 2\pi \alpha ^{\prime }\right)
^{2}\gamma ^{00}\mathbf{E}^{2}+\left( 2\pi \alpha ^{\prime }\right) ^{2}%
\mathbf{B}^{2}\right) \det \gamma -\left( 2\pi \alpha ^{\prime }\right)
^{4}\left( \mathbf{E}\cdot \mathbf{B}\right) ^{2}},
\end{equation*}%
where $\mathbf{E}^{2}=\gamma ^{mn}E_{m}E_{n}$ and $\mathbf{B}^{2}=\gamma
^{mn}B_{m}B_{n}$. We also redefine the tension $T_{3}$ by the dilaton field $%
\phi $ in the following way:%
\begin{equation*}
T_{3}e^{-\phi }\rightarrow T_{3}.
\end{equation*}%
In this redefinition is hidden an assumption that $\phi $ is constant on the
worldvolume. The metric $\gamma $ is induced by the backgrounds given by the
supergravity solutions, e.g. [7]. One can notice that the Lagrangian can be
rewritten in the form:%
\begin{equation}
L=T_{3}\sqrt{-\left( 1+\left( 2\pi \alpha ^{\prime }\right) ^{2}\mathbf{B}%
^{2}\right) \det \gamma -\mathbf{E}^{T}M\mathbf{E}},  \tag{2.3}
\end{equation}%
where the entries of the matrix $M$\ are given by:%
\begin{equation}
M^{mn}=\left( 2\pi \alpha ^{\prime }\right) ^{2}\gamma ^{00}\gamma ^{mn}\det
\gamma +\left( 2\pi \alpha ^{\prime }\right) ^{4}B^{m}B^{n}  \tag{2.4}
\end{equation}%
and $B^{m}=\gamma ^{mn}B_{n}$.

The canonical coordinates for the embedding $X$\ and the gauge field $A$\
are:%
\begin{equation*}
\left( X^{M},P_{M}\right) ,\left( \Pi ^{\alpha },A_{\alpha }\right) .
\end{equation*}%
The canonical momenta are given by:%
\begin{equation}
P_{M}=\frac{\partial L}{\partial \left( \partial _{0}X^{M}\right) }=-\frac{%
T_{3}}{2}\sqrt{-\det \left( \gamma +\mathcal{F}\right) }\left( G^{\alpha
0}+G^{0\alpha }\right) \partial _{\alpha }X^{N}g_{MN},  \tag{2.5}
\end{equation}%
\begin{equation}
\Pi ^{m}=\frac{\partial L}{\partial \left( \partial _{0}A_{m}\right) }=-%
\frac{2\pi \alpha ^{\prime }T_{3}}{2}\sqrt{-\det \left( \gamma +\mathcal{F}%
\right) }\left( G^{m0}-G^{0m}\right) ,  \tag{2.6}
\end{equation}%
\begin{equation}
\Pi ^{0}=\frac{\partial L}{\partial \left( \overset{\cdot }{A_{0}}\right) }%
=0,  \tag{2.7}
\end{equation}%
where:%
\begin{equation*}
G^{\alpha \beta }=\left( G^{-1}\right) ^{\alpha \beta }=\left[ \left( \gamma
+\mathcal{F}\right) ^{-1}\right] ^{\alpha \beta }.
\end{equation*}%
We define the following matrices:%
\begin{equation}
\mathcal{P}_{M}\mathcal{=}\left( \mathcal{P}_{M}^{\alpha }\right)
=G^{-T}e_{M}+G^{-1}e_{M},  \tag{2.8}
\end{equation}%
\begin{equation}
\mathcal{E=}\left( \mathcal{E}^{\alpha \beta }\right) \mathcal{=}%
G^{-1}-G^{-T},  \tag{2.9}
\end{equation}%
where:%
\begin{eqnarray*}
e_{M} &=&g_{MN}e^{N}, \\
e^{N} &=&\left( e_{\alpha }^{N}\right) =\left( \partial _{\alpha
}X^{N}\right) .
\end{eqnarray*}%
One can observe that the matrices $\mathcal{P}$\ and $\mathcal{E}$\ obey the
relation:%
\begin{equation}
\mathcal{P}_{M}e^{M}+\mathcal{EF=}2I,  \tag{2.10}
\end{equation}%
in the worldvolume coordinates this relation has the form:%
\begin{equation}
\mathcal{P}_{M}^{\alpha }e_{\beta }^{M}+\mathcal{E}^{\alpha \gamma }\mathcal{%
F}_{\gamma \beta }\mathcal{=}2\delta _{\beta }^{\alpha }.  \tag{2.10a}
\end{equation}%
The square of $\mathcal{P}_{M}$\ is:%
\begin{equation}
g^{MN}\mathcal{P}_{M}\mathcal{P}_{N}=\mathcal{E}\gamma \mathcal{E+}%
4G^{-1}+2\left( G^{-1}\mathcal{F}G^{-T}+G^{-T}\mathcal{F}G^{-1}\right) . 
\tag{2.11}
\end{equation}%
From the relation (2.10a) the following formulas for $\alpha =0$ and $\beta
=0$ have been obtained, respectively:%
\begin{equation}
2\pi \alpha ^{\prime }P_{M}e_{\beta }^{M}+\Pi ^{m}\mathcal{F}_{\beta m}=2\pi
\alpha ^{\prime }T_{3}\sqrt{-\det \left( \gamma +\mathcal{F}\right) }\delta
_{\beta }^{0},  \tag{2.12}
\end{equation}%
\begin{equation}
2\pi \alpha ^{\prime }P_{M}\partial _{0}X^{M}+\Pi ^{m}\mathcal{F}_{0m}=2\pi
\alpha ^{\prime }T_{3}\sqrt{-\det \left( \gamma +\mathcal{F}\right) }, 
\tag{2.13}
\end{equation}%
where $P_{M}$ and $\Pi ^{m}$ are related to $\mathcal{P}_{M}$ and $\mathcal{E%
}$ as follows:%
\begin{equation*}
P_{M}=T_{3}\sqrt{-\det \left( \gamma +\mathcal{F}\right) }\mathcal{P}%
_{M}^{0},
\end{equation*}%
\begin{equation*}
\Pi ^{m}=\frac{2\pi \alpha ^{\prime }T_{3}}{2}\sqrt{-\det \left( \gamma +%
\mathcal{F}\right) }\mathcal{E}^{0m}.
\end{equation*}%
For $\beta =m$\ one obtains the worldspace diffeomorphism constraint [8]:%
\begin{equation}
2\pi \alpha ^{\prime }P_{M}\partial _{m}X^{M}+\Pi ^{n}\mathcal{F}_{mn}=0. 
\tag{2.14}
\end{equation}%
There are also two other constraints [8]:

\begin{itemize}
\item the Hamiltonian constraint (which follows from (2.11)):%
\begin{equation*}
P_{M}P_{N}g^{MN}+\Pi ^{m}\Pi ^{n}\gamma _{mn}+T_{3}^{2}\det \left[ \left(
\gamma +\mathcal{F}\right) _{mn}\right] =0,
\end{equation*}

\item the Gauss law:%
\begin{equation*}
\partial _{m}\Pi ^{m}=0.
\end{equation*}
\end{itemize}

The Hamiltonian constraint was considered in the static embedding $X$ for
different configurations in [9].

Let us assume that the embedding $X$\ is not static and has the form:%
\begin{equation}
X\left( \xi \right) =\left( \xi ^{0},\xi ^{m},X^{a}\left( \xi ^{0},\xi
^{m}\right) \right)  \tag{2.15}
\end{equation}%
and the metric $g_{MN}$ is "diagonal":%
\begin{equation*}
\left( g_{MN}\right) =\left( 
\begin{array}{ccc}
g_{00} &  &  \\ 
& \left( g_{mn}\right) &  \\ 
&  & \left( g_{ab}\right)%
\end{array}%
\right)
\end{equation*}%
with the signature $\left( -1,+1,...,+1\right) .$ Thus $-g_{00}\geq 0$. For
the embedding$\ X$\ the induced metric $\gamma _{\alpha \beta }$\ takes the
form:%
\begin{eqnarray}
\gamma _{00} &=&g_{00}+g_{ab}\overset{\cdot }{X^{a}}\overset{\cdot }{X^{b}},
\notag \\
\gamma _{0m} &=&g_{ab}\overset{\cdot }{X^{a}}\partial _{m}X^{b},  \notag \\
\gamma _{m0} &=&g_{ab}\partial _{m}X^{a}\overset{\cdot }{X^{b}},  \notag \\
\gamma _{mn} &=&g_{mn}+g_{ab}\partial _{m}X^{a}\partial _{m}X^{b}. 
\TCItag{2.16}
\end{eqnarray}%
We restrict ourselves to a homogenous case: $\partial _{m}X^{a}=0$. Thus:%
\begin{equation*}
\det \left( \gamma _{\alpha \beta }\right) =\gamma _{00}\det \left( \gamma
_{mn}\right)
\end{equation*}%
and the matrix $M$ takes the form:%
\begin{equation*}
M^{mn}=\left( 2\pi \alpha ^{\prime }\right) ^{2}\gamma ^{mn}\det \left(
\gamma _{rp}\right) +\left( 2\pi \alpha ^{\prime }\right) ^{4}B^{m}B^{n}.
\end{equation*}%
The Lagrangian in this case takes the form:%
\begin{equation*}
L=T_{3}\sqrt{-\left( g_{00}+g_{ab}\overset{\cdot }{X^{a}}\overset{\cdot }{%
X^{b}}\right) \left( 1+\left( 2\pi \alpha ^{\prime }\right) ^{2}\mathbf{B}%
^{2}\right) \det \left( \gamma _{mn}\right) -\mathbf{E}M\mathbf{E}}.
\end{equation*}%
Note that $-\gamma _{00}>0$. The momenta $P_{a}$ transverse to the
worldvolume and the momenta $\Pi ^{m}$ have the form:%
\begin{equation}
P_{a}=-\frac{T_{3}^{2}\det \left( \gamma _{mn}\right) }{L}\left( 1+\left(
2\pi \alpha ^{\prime }\right) ^{2}\mathbf{B}^{2}\right) \overset{\cdot }{%
X^{b}}g_{ab}  \tag{2.17}
\end{equation}%
and%
\begin{equation}
\Pi ^{m}=-\frac{T_{3}^{2}}{L}M^{mn}E_{n},  \tag{2.18}
\end{equation}%
respectively. The tangent momentum to the worldvolume $P_{m}$ is obtained
from the diffeomorphism constraint (2.14):%
\begin{equation}
P_{m}=-\Pi ^{n}\mathcal{F}_{mn},  \tag{2.19}
\end{equation}%
and is expressed by the the Poynting vector $S_{m}=\varepsilon
_{mnp}E^{n}B^{p}$ on the worldvolume: 
\begin{equation*}
P_{m}=-\frac{T_{3}^{2}\left( 2\pi \alpha ^{\prime }\right) ^{3}\det \left(
\gamma _{rs}\right) }{L}S_{m},
\end{equation*}%
where $E^{n}=\gamma ^{nm}E_{m}$. The momentum $P_{M}$ has the form:%
\begin{equation}
P_{M}=\left( \mathcal{H},-\Pi ^{n}\mathcal{F}_{mn},P_{a}\right) ,  \tag{2.18}
\end{equation}%
where $\mathcal{H}$ is the energy density and $P_{a}$ is given by (2.17).
The square of $P_{M}$ is:%
\begin{equation*}
P_{M}P^{M}=g^{00}\mathcal{H}^{2}+g^{mn}P_{m}P_{n}+g^{ab}P_{a}P_{b},
\end{equation*}%
where:%
\begin{equation}
g^{mn}P_{m}P_{n}=\frac{\left( 2\pi \alpha ^{\prime }\right) ^{6}T_{3}^{4}}{%
L^{2}}\left[ \mathbf{E}\times \mathbf{B}\right] ^{2}\det {}^{2}\left( \gamma
_{mn}\right) ,  \tag{2.19}
\end{equation}%
\begin{equation}
g^{ab}P_{a}P_{b}=\frac{T_{3}^{4}}{L^{2}}\left( 1+\left( 2\pi \alpha ^{\prime
}\right) ^{2}\mathbf{B}^{2}\right) ^{2}\overset{\cdot }{X^{2}}\det
{}^{2}\left( \gamma _{mn}\right)  \tag{2.20}
\end{equation}%
and $\overset{\cdot }{X^{2}}=g_{ab}\overset{\cdot }{X^{a}}\overset{\cdot }{%
X^{b}}$ , the vector product is defined as:%
\begin{equation*}
\left( \mathbf{E}\times \mathbf{B}\right) _{m}=\varepsilon _{mnp}E^{n}B^{p}.
\end{equation*}%
The Hamiltonian constraint takes the form:%
\begin{equation*}
P_{M}P^{M}+\Pi ^{m}\Pi ^{n}\gamma _{mn}+T_{3}^{2}\left( 1-\frac{1}{2}%
\mathcal{F}_{mn}\mathcal{F}^{nm}\right) \det \left( \gamma _{pr}\right) =0,
\end{equation*}%
where:%
\begin{eqnarray*}
\Pi ^{m}\Pi ^{n}\gamma _{mn} &=&\frac{\left( 2\pi \alpha ^{\prime }\right)
^{4}T_{3}^{4}}{L^{2}}\left[ \mathbf{E}^{2}\det {}^{2}\left( \gamma
_{mn}\right) +2\left( 2\pi \alpha ^{\prime }\right) ^{2}\left( \mathbf{E}%
\cdot \mathbf{B}\right) ^{2}\det \left( \gamma _{mn}\right) +\left( 2\pi
\alpha ^{\prime }\right) ^{4}\mathbf{B}^{2}\left( \mathbf{E}\cdot \mathbf{B}%
\right) ^{2}\right] = \\
&&\frac{\left( 2\pi \alpha ^{\prime }\right) ^{4}T_{3}^{4}}{L^{2}}\left[ 
\mathbf{E}\det \left( \gamma _{mn}\right) +\left( 2\pi \alpha ^{\prime
}\right) ^{2}\left( \mathbf{E}\cdot \mathbf{B}\right) \mathbf{B}\right] ^{2},
\end{eqnarray*}%
\begin{equation*}
\frac{1}{2}\mathcal{F}_{mn}\mathcal{F}^{nm}=-\left( 2\pi \alpha ^{\prime
}\right) ^{2}\mathbf{B}^{2}.
\end{equation*}%
Thus the square of the energy density is:%
\begin{gather}
-g^{00}\mathcal{H}^{2}L^{2}=\left( 2\pi \alpha ^{\prime }\right)
^{2}T_{3}^{4}\left[ \mathbf{E}\times \mathbf{B}\right] ^{2}\det {}^{2}\left(
\gamma _{mn}\right) +\left( 2\pi \alpha ^{\prime }\right) ^{2}T_{3}^{4}\left[
\mathbf{E}\det \left( \gamma _{mn}\right) +\left( 2\pi \alpha ^{\prime
}\right) ^{2}\mathbf{B}\left( \mathbf{E}\cdot \mathbf{B}\right) \right] ^{2}+
\notag \\
T_{3}^{4}\left( 1+\left( 2\pi \alpha ^{\prime }\right) ^{2}\mathbf{B}%
^{2}\right) ^{2}\overset{\cdot }{X^{2}}\det {}^{2}\left( \gamma _{mn}\right)
+T_{3}^{2}L^{2}\left( 1+\left( 2\pi \alpha ^{\prime }\right) ^{2}\mathbf{B}%
^{2}\right) \det \left( \gamma _{mn}\right) .  \tag{2.21}
\end{gather}%
For $\mathbf{B=E}=0$ the Lagrangian (2.3) has the form:%
\begin{equation*}
L=T_{3}\sqrt{-\left( g_{00}+g_{ab}\overset{\cdot }{X^{a}}\overset{\cdot }{%
X^{b}}\right) \det \left( \gamma _{mn}\right) }.
\end{equation*}%
Hence one can obtain:%
\begin{equation}
\mathcal{H}^{2}=T_{3}^{2}\frac{-g_{00}\det \left( \gamma _{mn}\right) }{%
1+g^{00}\overset{\cdot }{X^{2}}},  \tag{2.22}
\end{equation}%
and:%
\begin{equation}
P_{a}=T_{3}\frac{\overset{\cdot }{X^{b}}g_{ab}}{\sqrt{1+g^{00}\overset{\cdot 
}{X^{2}}}}\sqrt{\frac{\det \left( \gamma _{mn}\right) }{-g_{00}}}. 
\tag{2.23}
\end{equation}%
Note that $|g_{00}|\geq \overset{\cdot }{X^{2}}$. The Hamiltonian constraint
has the form:%
\begin{equation*}
-g^{00}\mathcal{H}^{2}-\mathbf{P}^{2}=T_{3}^{2}\det \left( \gamma
_{mn}\right) .
\end{equation*}%
where $\mathbf{P}^{2}=g^{ab}P_{a}P_{b}$ . This last equation has the same
structure as the equation of the motion of a relativistic particle with the
mass $m_{0}=T_{3}\sqrt{\det \left( \gamma _{mn}\right) }$ . In the case of
D0-brane (D-particle) moving in the background with the metric $g_{MN}$ the
formulas (2.22) and (2.23) give the energy and momentum of the D-particle:%
\begin{equation}
\mathcal{H=}\frac{T_{0}\sqrt{-g_{00}}}{1+g^{00}\overset{\cdot }{X^{2}}}, 
\tag{2.24}
\end{equation}%
\begin{equation}
P_{a}=-\frac{T_{0}\sqrt{-g^{00}}\overset{\cdot }{X^{b}}g_{ab}}{\sqrt{1+g^{00}%
\overset{\cdot }{X^{2}}}}.  \tag{2.25}
\end{equation}%
The mass of this D-particle is $T_{0}$.

The energy density as the function of the momenta has the form:%
\begin{equation*}
\mathcal{H}=\sqrt{-g_{00}\left( T_{3}^{2}\left( 1+\left( 2\pi \alpha
^{\prime }\right) ^{2}\mathbf{B}^{2}\right) \det \left( \gamma _{mn}\right)
+g^{ab}P_{a}P_{b}+g^{mn}P_{m}P_{n}+\Pi ^{m}\Pi ^{n}\gamma _{mn}\right) }.
\end{equation*}%
Note that:%
\begin{equation*}
g^{mn}P_{m}P_{n}=\left( 2\pi \alpha ^{\prime }\right) ^{2}\Pi ^{m}\left[ 
\mathbf{B}^{2}\gamma _{mn}-B_{m}B_{n}\right] \Pi ^{n},
\end{equation*}%
since in this embedding $g_{mn}=\gamma _{mn}$. Hence the energy density is
given by:%
\begin{equation}
\mathcal{H}=\sqrt{-g_{00}\left[ T_{3}^{2}\left( 1+\left( 2\pi \alpha
^{\prime }\right) ^{2}\mathbf{B}^{2}\right) \det \left( \gamma _{mn}\right)
+g^{ab}P_{a}P_{b}+\Pi ^{m}\Pi ^{n}W_{mn}\right] },  \tag{2.26}
\end{equation}%
where:%
\begin{equation*}
W_{mn}=\left( 1+\left( 2\pi \alpha ^{\prime }\right) ^{2}\mathbf{B}%
^{2}\right) \gamma _{mn}-\left( 2\pi \alpha ^{\prime }\right) ^{2}B_{m}B_{n}.
\end{equation*}%
The energy density is the monotonically increasing function of the momenta.
So it is bounded from bottom by:%
\begin{equation*}
\mathcal{H}\geq T_{3}\sqrt{-g_{00}\left( 1+\left( 2\pi \alpha ^{\prime
}\right) ^{2}\mathbf{B}^{2}\right) \det \left( \gamma _{mn}\right) }.
\end{equation*}%
The equation (2.21) can also be expressed as a sum of the squares:%
\begin{align}
-g^{00}\mathcal{H}^{2}& =\frac{\left( 2\pi \alpha ^{\prime }\right)
^{2}T_{3}^{4}}{L^{2}}\left[ \mathbf{E}\times \mathbf{B}\det \left( \gamma
_{mn}\right) +\mathbf{E}\det \left( \gamma _{mn}\right) +\mathbf{B}\left( 
\mathbf{E}\cdot \mathbf{B}\right) \right] ^{2}+  \notag \\
& \frac{T_{3}^{2}\left( 1+\left( 2\pi \alpha ^{\prime }\right) ^{2}\mathbf{B}%
^{2}\right) }{2L^{2}}\left[ T_{3}\sqrt{1+\left( 2\pi \alpha ^{\prime
}\right) ^{2}\mathbf{B}^{2}}\left\vert \overset{\cdot }{X}\right\vert \sqrt{%
\det \left( \gamma _{mn}\right) }+L\right] ^{2}\det \left( \gamma
_{mn}\right) +  \notag \\
& \frac{T_{3}^{2}\left( 1+\left( 2\pi \alpha ^{\prime }\right) ^{2}\mathbf{B}%
^{2}\right) }{2L^{2}}\left[ T_{3}\sqrt{1+\left( 2\pi \alpha ^{\prime
}\right) ^{2}\mathbf{B}^{2}}\left\vert \overset{\cdot }{X}\right\vert \sqrt{%
\det \left( \gamma _{mn}\right) }-L\right] ^{2}\det \left( \gamma
_{mn}\right) .  \tag{2.27}
\end{align}%
One can deduce that the energy square is bounded by the following
configurations:%
\begin{gather}
-g^{00}\mathcal{H}^{2}\geq \frac{\left( 2\pi \alpha ^{\prime }\right)
^{2}T_{3}^{4}}{L^{2}}\left[ \mathbf{E}\times \mathbf{B}\det \left( \gamma
_{mn}\right) +\mathbf{E}\det \left( \gamma _{mn}\right) +\mathbf{B}\left( 
\mathbf{E}\cdot \mathbf{B}\right) \right] ^{2}+  \notag \\
\frac{T_{3}^{2}\left( 1+\left( 2\pi \alpha ^{\prime }\right) ^{2}\mathbf{B}%
^{2}\right) }{2L^{2}}\left[ T_{3}\sqrt{1+\left( 2\pi \alpha ^{\prime
}\right) ^{2}\mathbf{B}^{2}}\left\vert \overset{\cdot }{X}\right\vert \sqrt{%
\det \left( \gamma _{mn}\right) }+L\right] ^{2}\det \left( \gamma
_{mn}\right) .  \tag{2.28}
\end{gather}%
One obtains the equality when:%
\begin{equation}
T_{3}\sqrt{1+\left( 2\pi \alpha ^{\prime }\right) ^{2}\mathbf{B}^{2}}%
\left\vert \overset{\cdot }{X}\right\vert \sqrt{\det \left( \gamma
_{mn}\right) }=L.  \tag{2.29}
\end{equation}%
The inequality (2.28) is the BPS bound [9]. In the case when $\mathbf{E=B=0}$
the condition (2.29) gives:%
\begin{equation*}
2\overset{\cdot }{X^{2}}\det \left( \gamma _{mn}\right) =-g_{00}.
\end{equation*}%
For the static configuration $\overset{\cdot }{X^{a}}=0$ one obtains:%
\begin{equation*}
-g_{00}\left( 1+\left( 2\pi \alpha ^{\prime }\right) ^{2}\mathbf{B}%
^{2}\right) \det \left( \gamma _{mn}\right) =\mathbf{E}M\mathbf{E}
\end{equation*}

The BPS configuration for $\overset{\cdot }{X^{a}}\neq 0$ has the energy:%
\begin{gather}
-g^{00}\mathcal{H}_{BPS}^{2}=T_{3}^{2}\frac{\left( 2\pi \alpha ^{\prime
}\right) ^{2}\left[ \mathbf{E}\times \mathbf{B}\det \left( \gamma
_{mn}\right) +\mathbf{E}\det \left( \gamma _{mn}\right) +\mathbf{B}\left( 
\mathbf{E}\cdot \mathbf{B}\right) \right] ^{2}}{\left( 1+\left( 2\pi \alpha
^{\prime }\right) ^{2}\mathbf{B}^{2}\right) \overset{\cdot }{X^{2}}\det
\left( \gamma _{mn}\right) }+  \notag \\
+2T_{3}^{2}\left( 1+\left( 2\pi \alpha ^{\prime }\right) ^{2}\mathbf{B}%
^{2}\right) ^{2}\det \left( \gamma _{mn}\right) .  \tag{2.30}
\end{gather}

The expression under the square root in (2.3) has to be positive in order to
gets the real Lagrangian. This condition puts the constraint on the allowed
configurations:%
\begin{equation}
-\left( 1+\left( 2\pi \alpha ^{\prime }\right) ^{2}\gamma ^{00}\mathbf{E}%
^{2}+\left( 2\pi \alpha ^{\prime }\right) ^{2}\mathbf{B}^{2}\right) \det
\gamma \geq \left( 2\pi \alpha ^{\prime }\right) ^{4}\left( \mathbf{E}\cdot 
\mathbf{B}\right) ^{2},  \tag{2.31}
\end{equation}%
since $\left( \mathbf{E}\cdot \mathbf{B}\right) ^{2}>0$. Integrating the
square root\ of (2.31) over the D3-brane $M$ one obtains:%
\begin{equation*}
\int_{M}\sqrt{-\left( 1+\left( 2\pi \alpha ^{\prime }\right) ^{2}\mathbf{B}%
^{2}+\left( 2\pi \alpha ^{\prime }\right) ^{2}\gamma ^{00}\mathbf{E}%
^{2}\right) \det \gamma }d^{4}x\geq \left( 2\pi \alpha ^{\prime }\right)
^{2}\int_{M}\mathbf{E}\cdot \mathbf{B}d^{4}x.
\end{equation*}%
The r.h.s. of the above inequality is expressed by the second Chern
character $ch_{2}\left( L\right) $ of the line bundle $L$ over $M$. This
second Chern character is expressed by the curvature form $F$ of the line
bundle as follows:%
\begin{equation*}
ch_{2}\left( L\right) =\frac{-1}{8\pi ^{2}}F\wedge F=\frac{1}{2\pi ^{2}}%
\mathbf{E}\cdot \mathbf{B}d^{4}x.
\end{equation*}%
Thus one obtains:%
\begin{equation}
\int_{M}\sqrt{-\left( 1+\left( 2\pi \alpha ^{\prime }\right) ^{2}\mathbf{B}%
^{2}+\left( 2\pi \alpha ^{\prime }\right) ^{2}\gamma ^{00}\mathbf{E}%
^{2}\right) \det \gamma }d^{4}x\geq 2\pi ^{2}\left( 2\pi \alpha ^{\prime
}\right) ^{2}\int_{M}ch_{2}\left( L\right) .  \tag{2.32}
\end{equation}%
For the embedding under consideration the formula (2.31) takes the form:%
\begin{equation}
\left( -\gamma _{00}-\left( 2\pi \alpha ^{\prime }\right) ^{2}\mathbf{E}%
^{2}-\left( 2\pi \alpha ^{\prime }\right) ^{2}\gamma _{00}\mathbf{B}%
^{2}\right) \det \left( \gamma _{mn}\right) \geq \left( 2\pi \alpha ^{\prime
}\right) ^{4}\left( \mathbf{E}\cdot \mathbf{B}\right) ^{2}.  \tag{2.33}
\end{equation}%
One can notice that:%
\begin{equation}
-\gamma _{00}-\left( 2\pi \alpha ^{\prime }\right) ^{2}\mathbf{E}^{2}-\left(
2\pi \alpha ^{\prime }\right) ^{2}\gamma _{00}\mathbf{B}^{2}\geq 0. 
\tag{2.34}
\end{equation}%
This condition leads to the relation:%
\begin{equation*}
-\gamma _{00}\geq \left( 2\pi \alpha ^{\prime }\right) ^{2}\mathbf{E}%
^{2}+\left( 2\pi \alpha ^{\prime }\right) ^{2}\gamma _{00}\mathbf{B}^{2}
\end{equation*}
($-\gamma _{00}>0)$. Thus in the induced metric $\gamma $ with the signature 
$\left( -,+,+,+\right) $ we get a bound on the allowed magnetic and electric
fields:%
\begin{equation*}
|\gamma _{00}|\geq \left( 2\pi \alpha ^{\prime }\right) ^{2}\mathbf{E}%
^{2}-\left( 2\pi \alpha ^{\prime }\right) ^{2}|\gamma _{00}|\mathbf{B}^{2}.
\end{equation*}%
This relation is in agreement with the\ result of [10] which says that the
electric field has the maximal value. In the case when the last inequality
is saturated, it means that fields $\mathbf{E}$ and $\mathbf{B}$ are
maximal. Thus one obtains that $\mathbf{E}\cdot \mathbf{B=0,}$ so the DBI
Lagrangian vanishes.

\section{Evolution of the D3-brane}

In this section we consider a D3-brane embedded in the\ non-static and
homogenous way\ in the background given by the supergravity solutions. Let
us deform the time component of the metric $\gamma $ in such a way that\ it
becomes original for $\mathbf{E=B}=0$. The simplest deformation\ which
fulfils the above conditions is obtained from (2.34). $\ \ $

The condition (2.34) can be expressed by the function $V$ as follows:%
\begin{equation}
V\geq 0,  \tag{3.1}
\end{equation}%
where:%
\begin{equation}
V=-\gamma _{00}\left( 1+\left( 2\pi \alpha ^{\prime }\right) ^{2}\mathbf{B}%
^{2}\right) -\left( 2\pi \alpha ^{\prime }\right) ^{2}\mathbf{E}^{2}. 
\tag{3.2}
\end{equation}%
The condition (3.1) agrees with the signature of the induced metric $\left(
-,+,+,+\right) .$ Thus the deformed\emph{\ }metric $dl^{\prime 2}$ on the
brane looks like:%
\begin{equation}
dl^{\prime 2}=-V\left( d\xi ^{0}\right) ^{2}+\gamma _{mn}d\xi ^{m}d\xi ^{n}.
\tag{3.3}
\end{equation}%
The condition that $V=0$ determines a certain region on which the metric $%
dl^{\prime 2}$ is degenerated. These configurations, for which $V=0,$
correspond to the vanishing of the Lagrangian $L$.

Let us consider the background metric given by:%
\begin{equation}
ds^{2}=-\lambda _{0}dt^{2}+\lambda _{1}\sum_{i=1}^{\widetilde{d}%
-1}dX_{i}^{2}+\lambda _{2}dr^{2}+r^{2}\lambda _{3}d\Omega _{d+1}.  \tag{3.4}
\end{equation}%
This metric describes a $(d+2)$-brane which is wrapped on $S^{d+1}$. For the
embedding 
\begin{equation}
X^{M}\left( \xi ^{0},\xi ^{m}\right) =\left( \xi ^{0},\xi ^{m},X^{a}\left(
\xi ^{0}\right) \right)  \tag{3.5}
\end{equation}%
(where $a=4,...,9)$ the induced $dl^{2}$ metric on $M$ has the form:%
\begin{eqnarray}
\gamma _{00} &=&-\lambda _{0}+\lambda _{1}\sum_{i=4}^{\widetilde{d}-1}%
\overset{\cdot }{X}_{i}^{2}+\lambda _{2}\overset{\cdot }{r}^{2}+r^{2}\lambda
_{3}\overset{\cdot }{\mathbf{\varphi }}^{2},  \TCItag{3.6} \\
\gamma _{mn} &=&\lambda _{1}\delta _{mn},\text{for }3\leq \widetilde{d}-1. 
\TCItag{3.7}
\end{eqnarray}%
In ten dimensions $\widetilde{d}-1=9-d$ and $\gamma _{m0}=0$,\ where:%
\begin{equation}
\overset{\cdot }{\mathbf{\varphi }}^{2}=h_{rs}\overset{\cdot }{\varphi }^{r}%
\overset{\cdot }{\varphi }^{s},  \tag{3.8}
\end{equation}%
and $h_{rs}=h_{rs}\left( \varphi \right) $ ($r,s=1,...,d+1$)\ is the metric
on $S^{d+1}$. Thus the deformed metric is:%
\begin{eqnarray}
dl^{\prime }{}^{2} &=&-\left( \left( \lambda _{0}-\lambda _{1}\sum_{i=4}^{%
\widetilde{d}-1}\overset{\cdot }{X}_{i}^{2}-\lambda _{2}\overset{\cdot }{r}%
^{2}-r^{2}\lambda _{3}\overset{\cdot }{\mathbf{\varphi }}^{2}\right) \left(
1+\left( 2\pi \alpha ^{\prime }\right) ^{2}\mathbf{B}^{2}\right) -\left(
2\pi \alpha ^{\prime }\right) ^{2}\mathbf{E}^{2}\right) \left( d\xi
^{0}\right) ^{2}+  \notag \\
&&\lambda _{1}d\xi _{n}d\xi ^{n}.  \TCItag{3.9}
\end{eqnarray}%
Using spherical coordinates $\left( \rho ,\theta ,\psi \right) $ on $M$ the
above metric assumes the form:

\begin{eqnarray}
dl^{\prime }{}^{2} &=&-\left( \left( \lambda _{0}-\lambda _{1}\sum_{i=4}^{%
\widetilde{d}-1}\overset{\cdot }{X}_{i}^{2}-\lambda _{2}\overset{\cdot }{r}%
^{2}-r^{2}\lambda _{3}\overset{\cdot }{\mathbf{\varphi }}^{2}\right) \left(
1+\left( 2\pi \alpha ^{\prime }\right) ^{2}\mathbf{B}^{2}\right) -\left(
2\pi \alpha ^{\prime }\right) ^{2}\mathbf{E}^{2}\right) \left( d\xi
^{0}\right) ^{2}+  \notag \\
&&\lambda _{1}d\rho ^{2}+\rho ^{2}\lambda _{1}d\Omega _{2}.  \TCItag{3.10}
\end{eqnarray}%
Let us now compare this metric with the Reissner-Nordstr\"{o}m-like metric
which describes a charged black hole. The standard form of the metric
describing a Reissner-Nordstr\"{o}m-like black hole in four dimensions is
the following:%
\begin{equation}
ds^{2}=-f\left( r\right) dt^{2}+f^{-1}\left( r\right) dr^{2}+r^{2}d\Omega
_{2}  \tag{3.11}
\end{equation}%
where $f$\ is equal to zero for two values of $r$. These zeros are ordered
in such a way: $r_{+}>r_{-}$, and $r_{+}$\ defines an event horizon. In the
case when $r_{+}=r_{-}$\ the black hole is extremal. In order to obtain the
Reissner-Nordstr\"{o}m-like black hole for the metric $dl^{\prime }$ the
following constraint should be put on the metric components:%
\begin{equation}
\left( \left( \lambda _{0}-\lambda _{1}\sum_{i=4}^{\widetilde{d}-1}\overset{%
\cdot }{X}_{i}^{2}-\lambda _{2}\overset{\cdot }{r}^{2}-r^{2}\lambda _{3}%
\overset{\cdot }{\mathbf{\varphi }}^{2}\right) \left( 1+\left( 2\pi \alpha
^{\prime }\right) ^{2}\mathbf{B}^{2}\right) -\left( 2\pi \alpha ^{\prime
}\right) ^{2}\mathbf{E}^{2}\right) \lambda _{1}=1.  \tag{3.12}
\end{equation}%
In the case when $\overset{\cdot }{r}=\overset{\cdot }{\mathbf{\varphi }}=0$
the above condition assumes the form\emph{:}%
\begin{equation*}
\lambda _{1}^{2}\left( 1+\left( 2\pi \alpha ^{\prime }\right) ^{2}\mathbf{B}%
^{2}\right) \sum_{i=4}^{\widetilde{d}-1}\overset{\cdot }{X}_{i}^{2}-\lambda
_{1}\left[ \lambda _{0}\left( 1+\left( 2\pi \alpha ^{\prime }\right) ^{2}%
\mathbf{B}^{2}\right) -\left( 2\pi \alpha ^{\prime }\right) ^{2}\mathbf{E}%
^{2}\right] +1=0.
\end{equation*}%
The solutions for this equations with respect to $\lambda _{1}$\ are as
follows:%
\begin{gather}
\lambda _{1\left( \pm \right) }=\frac{\lambda _{0}\left( 1+\left( 2\pi
\alpha ^{\prime }\right) ^{2}\mathbf{B}^{2}\right) -\left( 2\pi \alpha
^{\prime }\right) ^{2}\mathbf{E}^{2}}{2\left( 1+\left( 2\pi \alpha ^{\prime
}\right) ^{2}\mathbf{B}^{2}\right) \sum_{i=4}^{\widetilde{d}-1}\overset{%
\cdot }{X}_{i}^{2}}  \notag \\
\pm \frac{\sqrt{\left[ \lambda _{0}\left( 1+\left( 2\pi \alpha ^{\prime
}\right) ^{2}\mathbf{B}^{2}\right) -\left( 2\pi \alpha ^{\prime }\right) ^{2}%
\mathbf{E}^{2}\right] ^{2}-4\left( 1+\left( 2\pi \alpha ^{\prime }\right)
^{2}\mathbf{B}^{2}\right) \sum_{i=4}^{\widetilde{d}-1}\overset{\cdot }{X}%
_{i}^{2}}}{2\left( 1+\left( 2\pi \alpha ^{\prime }\right) ^{2}\mathbf{B}%
^{2}\right) \sum_{i=4}^{\widetilde{d}-1}\overset{\cdot }{X}_{i}^{2}}. 
\tag{3.13}
\end{gather}%
$\lambda _{1}$ is real if:%
\begin{equation}
D=\left[ \lambda _{0}\left( 1+\left( 2\pi \alpha ^{\prime }\right) ^{2}%
\mathbf{B}^{2}\right) -\left( 2\pi \alpha ^{\prime }\right) ^{2}\mathbf{E}%
^{2}\right] ^{2}-4\left( 1+\left( 2\pi \alpha ^{\prime }\right) ^{2}\mathbf{B%
}^{2}\right) \sum_{i=4}^{\widetilde{d}-1}\overset{\cdot }{X}_{i}^{2}\geq 0. 
\tag{3.14}
\end{equation}%
The only one positive solution of (3.13) exists when $D=0$. This condition
relates $\lambda _{0}$ to $\overset{\cdot }{X}_{i}$, $\mathbf{E}$ and $%
\mathbf{B}$:%
\begin{equation}
\left[ \lambda _{0}\left( 1+\left( 2\pi \alpha ^{\prime }\right) ^{2}\mathbf{%
B}^{2}\right) -\left( 2\pi \alpha ^{\prime }\right) ^{2}\mathbf{E}^{2}\right]
^{2}=4\left( 1+\left( 2\pi \alpha ^{\prime }\right) ^{2}\mathbf{B}%
^{2}\right) \sum_{i=4}^{\widetilde{d}-1}\overset{\cdot }{X}_{i}^{2}. 
\tag{3.15}
\end{equation}%
In this case the solution (3.13) is:%
\begin{equation}
\lambda _{1}=\frac{1}{\sqrt{\left( 1+\left( 2\pi \alpha ^{\prime }\right)
^{2}\mathbf{B}^{2}\right) }\sqrt{\sum_{i=4}^{\widetilde{d}-1}\overset{\cdot }%
{X}_{i}^{2}}}  \tag{3.16a}
\end{equation}%
and:%
\begin{equation}
\lambda _{0}=\frac{\left( 2\pi \alpha ^{\prime }\right) ^{2}\mathbf{E}%
^{2}\pm 2\sqrt{\left( 1+\left( 2\pi \alpha ^{\prime }\right) ^{2}\mathbf{B}%
^{2}\right) \sum_{i=4}^{\widetilde{d}-1}\overset{\cdot }{X}_{i}^{2}}}{%
1+\left( 2\pi \alpha ^{\prime }\right) ^{2}\mathbf{B}^{2}}.  \tag{3.16b}
\end{equation}%
Note that $\lambda _{0}>0$\ for all configurations if one chooses the sign $%
+ $\ in (3.16b). In the case when the sign $-$\ is chosen the allowed
configurations are restricted by the following condition:%
\begin{equation*}
\frac{\left( 2\pi \alpha ^{\prime }\right) ^{2}\mathbf{E}^{2}}{2\sqrt{%
1+\left( 2\pi \alpha ^{\prime }\right) ^{2}\mathbf{B}^{2}}}\geq \sqrt{%
\sum_{i=4}^{\widetilde{d}-1}\overset{\cdot }{X}_{i}^{2}}.
\end{equation*}%
Thus the metric $dl^{\prime }$ has the form:%
\begin{equation}
dl^{2\prime }=-\sqrt{\left( 1+\left( 2\pi \alpha ^{\prime }\right) ^{2}%
\mathbf{B}^{2}\right) }\sqrt{\sum_{i=4}^{\widetilde{d}-1}\overset{\cdot }{X}%
_{i}^{2}}\left( d\xi ^{0}\right) ^{2}+\frac{1}{\sqrt{\left( 1+\left( 2\pi
\alpha ^{\prime }\right) ^{2}\mathbf{B}^{2}\right) }\sqrt{\sum_{i=4}^{%
\widetilde{d}-1}\overset{\cdot }{X}_{i}^{2}}}\left( d\rho ^{2}+\rho
^{2}d\Omega _{2}\right) .  \tag{3.17}  \label{46}
\end{equation}%
In the static case, i.e. $\overset{\cdot }{X}_{i}=\overset{\cdot }{r}=%
\overset{\cdot }{\mathbf{\varphi }}=0,$ the condition (3.12) gives:%
\begin{equation}
\left( \lambda _{0}\left( 1+\left( 2\pi \alpha ^{\prime }\right) ^{2}\mathbf{%
B}^{2}\right) -\left( 2\pi \alpha ^{\prime }\right) ^{2}\mathbf{E}%
^{2}\right) \lambda _{1}=1.  \tag{3.18}  \label{47}
\end{equation}%
Thus:%
\begin{eqnarray}
dl^{2\prime } &=&-\left( \lambda _{0}\left( 1+\left( 2\pi \alpha ^{\prime
}\right) ^{2}\mathbf{B}^{2}\right) -\left( 2\pi \alpha ^{\prime }\right) ^{2}%
\mathbf{E}^{2}\right) \left( d\xi ^{0}\right) ^{2}+  \notag \\
&&\frac{1}{\lambda _{0}\left( 1+\left( 2\pi \alpha ^{\prime }\right) ^{2}%
\mathbf{B}^{2}\right) -\left( 2\pi \alpha ^{\prime }\right) ^{2}\mathbf{E}%
^{2}}\left( d\rho ^{2}+\rho ^{2}d\Omega _{2}\right) .  \TCItag{3.19}
\label{48}
\end{eqnarray}%
This metric describes the spacetime with the magnetic and electric fields.
Moreover this spacetime has the event horizon which is given by the
vanishing of the Lagrangian since \TEXTsymbol{\vert}$\gamma _{00}|=|\lambda
_{0}|$.

In the metric (3.17) we make the change of the variable $\xi _{0}$ assuming
that the magnetic field $B$\ is constant on $M$:%
\begin{equation}
\tau \left( t\right) =\int^{t}F\left( t^{\prime }\right) dt^{\prime }, 
\tag{3.20}  \label{49}
\end{equation}%
where $t=\xi _{0}$\ and:%
\begin{equation}
F\left( t\right) =\left[ \left( 1+\left( 2\pi \alpha ^{\prime }\right) ^{2}%
\mathbf{B}^{2}\right) \sum_{i=4}^{\widetilde{d}-1}\overset{\cdot }{X}%
_{i}^{2}\left( t\right) \right] ^{1/4}.  \tag{3.21}  \label{50}
\end{equation}%
In\ this\ new coordinate $\tau $ the metric (3.17) takes the form:%
\begin{equation}
dl^{2\prime }=-d\tau ^{2}+\left[ \frac{1}{F\left( f\left( \tau \right)
\right) }\right] ^{2}\left( d\rho ^{2}+\rho ^{2}d\Omega _{2}\right) , 
\tag{3.22}  \label{51}
\end{equation}%
where the function $f\left( \tau \right) $\ is the inverse function to the
function (3.20):%
\begin{equation*}
t=f\left( \tau \right) .
\end{equation*}%
In order to predict, how does the deformed metric (3.17) behave, let us
assume the following form of the function $F\left( t\right) $:%
\begin{equation}
F\left( t\right) =\Lambda ^{1/2}t^{\alpha /2},  \tag{3.23 }
\end{equation}%
where $\alpha $\ and $\Lambda $\ are constants. This function is related to
the Kaster-Traschen dynamic solutions [11] in the case when $\alpha =1$.
These solutions have been generalized to the branes in [12]\emph{.} The
metric (3.17) for this function takes the form: 
\begin{equation}
dl^{2\prime }=-\Lambda t^{\alpha /2}dt^{2}+\Lambda ^{-1}t^{-\alpha /2}d%
\mathbf{x}^{2}.  \tag{3.24}  \label{53}
\end{equation}%
In the coordinate $\tau $ related to $t$ by%
\begin{equation}
t\left( \tau \right) =\left[ \left( 1+\alpha /4\right) \sqrt{\Lambda }\tau %
\right] ^{4/\left( 4+\alpha \right) }  \tag{3.25}  \label{54}
\end{equation}%
(for $\alpha \neq -4$) the metric (3.24) becomes:%
\begin{equation}
dl^{2\prime }=-d\tau ^{2}+\Lambda ^{-1}\left[ \left( 1+\alpha /4\right) 
\sqrt{\Lambda }\right] ^{-2\alpha /\left( 4+\alpha \right) }\tau ^{-2\alpha
/\left( 4+\alpha \right) }d\mathbf{x}^{2}.  \tag{3.26}  \label{55}
\end{equation}%
In the case when $\alpha =-4$ the variables $t$ and $\tau $ are related with
each other as follows:%
\begin{equation}
t\left( \tau \right) =\exp \left( \tau /\sqrt{\Lambda }\right) .  \tag{3.27}
\label{56}
\end{equation}%
The metric (3.24) for $\alpha =-4$ is:%
\begin{equation}
dl^{2\prime }=-d\tau ^{2}+\Lambda ^{-1}\exp \left( -\tau /\sqrt{\Lambda }%
\right) d\mathbf{x}^{2}.  \tag{3.28}  \label{57}
\end{equation}%
If $\alpha \in \left( -\infty ,-4\right) \cup \left( 0,+\infty \right) $,
then\ $2\alpha /\left( 4+\alpha \right) >0$ and the metric (3.28) represents
the four-dimensional space-time being contracted from the phase with the
finite space intervals at $\tau =0$ to the phase with these intervals going
to zero. For $\alpha =0$ the metric is static. On the other hand the
expanding space-time is obtained for $\alpha \in \left( -4,0\right) $
(because $2\alpha /\left( 4+\alpha \right) <0$) starting from an initial
singularity. The Kasner metric is obtained for $\alpha =4/3$ ([12]). The
special case corresponds to $\alpha =-4$ with the metric (3.28). It is the
de Sitter metric being contracted from the maximal space distance $\Lambda
^{-1/2}$ for $\tau =0$. In this way one obtains a family of contracting and
expanding directions tangent to the D3-brane in the case when the constraint
(3.23) holds.

The distance $l$ in the transverse directions $X_{i}$ with respect to the
ambient metric (3.4) is given by:%
\begin{equation*}
l=\int \sqrt{\lambda _{1}}ds,
\end{equation*}%
where $s$ is a parameter on a curve in the $X_{i}$ directions. From (3.16)
and (3.23) one obtains:%
\begin{equation}
l=\left\{ 
\begin{array}{cc}
\frac{4t^{\left( 1-\alpha /4\right) }}{\sqrt{\Lambda }\left( 4-\alpha
\right) } & \text{for }\alpha \neq 4 \\ 
\frac{\ln t}{\sqrt{\Lambda }} & \text{for }\alpha =4%
\end{array}%
.\right.  \tag{3.29}  \label{58}
\end{equation}%
The distance $l$ expressed by $\tau $ has the form:%
\begin{equation}
l=\left\{ 
\begin{array}{c}
\begin{array}{cc}
\frac{4}{\sqrt{\Lambda }\left( 4-\alpha \right) }\left[ \left( 1+\alpha
/4\right) \sqrt{\Lambda }\right] ^{\frac{4-\alpha }{4+\alpha }}\tau ^{\frac{%
4-\alpha }{4+\alpha }} & \text{for }\alpha \neq -4,+4 \\ 
\frac{1}{2\sqrt{\Lambda }}\ln \left( 2\sqrt{\Lambda }\tau \right) & \text{%
for }\alpha =+4%
\end{array}
\\ 
\begin{array}{cc}
\frac{1}{2\sqrt{\Lambda }}\exp \left( 2\tau /\sqrt{\Lambda }\right) & \text{%
for }\alpha =-4%
\end{array}%
\end{array}%
.\right.  \tag{3.30}  \label{59}
\end{equation}%
The distance $l$ should be positive, so $4>\alpha $.\ The transverse
directions do expand for $\alpha \in \left( -4,+4\right) $ and contract for $%
\alpha \in \left( -\infty ,-4\right) $. The cases when $\alpha =-4$ and $%
\alpha =+4$ correspond to the expanding transverse directions. To summarize,
for some values of $\alpha $ the tangent directions contract while the
transverse directions expand:

\begin{itemize}
\item For $\alpha \in \left( -\infty ,-4\right) $ the tangent and transverse
directions contract.

\item For $\alpha \in \left( -4,0\right) $ the tangent and transverse
directions expand from the initial singularity.

\item For $\alpha \in (0,+4]$ the tangent directions contract while the
transverse directions expand.

\item For $\alpha =-4$ the tangent directions are described by the
contracting de Sitter metric with the maximal size $\Lambda ^{-1/2}$ whereas
the transverse directions expand. The minimal size of the transverse space
is $2^{-1}\Lambda ^{-1/2}.$

\item For $\alpha =0$ the tangent directions are static (they do not depend
on $\tau $), the transverse directions expand from an initial point which is
not singular.
\end{itemize}

\section{Conclusions}

The metric induced on the D3-brane has been deformed by adding the electric
and magnetic fields to $\gamma _{00}$. This new metric has been compared to
the Reissner-Nordstr\"{o}m-like metric. This comparison has been made
because we have expected the appearance of the singularities on the brane in
the case when the DBI Lagrangian vanishes. Vanishing the DBI Lagrangian can
be interpreted as a result of the strong coupling ( $T_{3}\rightarrow 0$)
[8]. In this case description of the brane by the DBI action is unvalid
since the DBI Lagrangian is obtained in the low energetic approximation.

A family of expanding and contracting branes has been obtained for the
embedding $X$ restricted by the Eq. (3.23). The DBI Lagrangian vanishes for $%
t=0$. This corresponds to a special state of the D3-brane. This state has
been interpreted as an initial singularity (in the case of expansion) or a
final singularity (in the case of contraction). Other solutions with the
vanishing DBI Lagrangian have been obtained which do not possess
singularities on the brane worldvolume. These solutions have been expressed
by the de Sitter metrics (Eq. (3.28)).

The behavior of the directions transverse to the D3-brane is given as a
function of time by Eq.(3.30). In this way a dynamical model of the
space-time with\ a D3-brane\ embedded in it has been obtained.

\section{References}

[1]A. Achucarro, J. M. Evans, P. K. Townsend, D. L. Wiltshire, \textit{Super
p-branes}, Phys. Lett. B198, 441 (1987) ; M. Aganagic, C. Popescu, J. H.
Schwarz, \textit{D-brane actions with local kappa symmetry}, Phys. Lett.
B393 (1997) pp.311-315, Phys. Lett. B495, 145 (1997).

[2] M. Marino, R. Minasian, G. W. Moore, A. Strominger, \textit{Nonlinear
Instantons from Supersymmetric p-Branes}, JHEP 0001 (2000) 005,
hep-th/9911206.

[3] R. Harvey, B. Lawson, \textit{Calibrated geometries}, Acta Math. 148
(1982) pp.57-157, D.Joyce, \textit{Lectures on Calabi-Yau Lagrangian geometry%
}, math.DG/0108088, P. Koerber, \textit{Stable D-branes, calibrations and
genaralized Calabi-Yau geometry}, hep-th/0506154

[4] C. P. Bachas, P. Bain, M. B. Green: \textit{Curvature terms in D-brane
actions and their M-theory origin}, JHEP 9905 (1999) 011, hep-th/9903210

[5] D. Mateos, S. Ng, P. K. Townsend, \textit{Tachyons, supertubes and
brane/anti-brane systems}, JHEP 03 (2002) 016.\ 

[6] M. R. Douglas, B. Fiol, and C. R\"{o}melsberger, \textit{Stability and
BPS Branes}, JHEP 0509 (2005) 006, hep-th/0002037.

[7] M. J. Duff, R. R. Khuri, J. X. Lu, \textit{String solitons}, Phys.Rept.
259 (1995) 213, hep-th/9412184 ; M. J. Duff, Supermembranes, hep-th/9611203

[8] U. Lindstr\"{o}m, R. von Unge, \textit{A Picture of D-branes at Strong
Coupling}, Phys.Lett. B403 (1997) 233-238, hep-th/9704051;\ R. Kallosh, 
\textit{Covariant quantization of D-branes}, Phys.Rev. D56 (1997) 3515-3522,
hep-th/9705056; K. Kamimura and M. Hatsuda, \textit{Canonical formulation of
IIB D-branes}, Nucl. Phys. B527 (1998) 381 ; hep-th/9712068

[9] J. P. Gauntlett, J. Gomis, P.K. Townsend, \textit{BPS bounds for
worldvolume branes}, JHEP 01 (1998) 003, hep-th/9711205.

[10] C. Bachas, M. Porrati, \textit{Pair creation of open strings in an
electric field}, Phys. Lett. B296 (1992) 77, hep-th/9209032 ; C. Bachas, 
\textit{D-brane dynamics}, Phys.Lett. B374 (1996) 37.

[11] D. Kastor, J. Traschen,\ \textit{Cosmological Multi-Black Hole
Solutions, }Phys.Rev. D47 (1993) 5370, hep-th/9212035

[12] G. W. Gibbons, H. Lu, C. N. Pope, \textit{Brane Worlds in Collision},
hep-th/0501117.

\end{document}